# On "Bit-Interleaved Coded Multiple Beamforming"

E. Akay[1], H. J. Park[2], E. Ayanoglu[2]

ABSTRACT

**The interleaver design criteria described in [1] should take into account all error patterns of interest.**

In [1], let $S$ be the number of streams employed and let $\alpha_s$ be the number of times the stream $s$ is employed with a corresponding channel bit equal to 1 in an error path. Then, the interleaver design criteria for single-carrier systems at the beginning of Section II of [1] should be as follows

For all error paths of interest,
1) the consecutive coded bits are mapped over different symbols,
2) $\alpha_s \geq 1$ for $1 \leq s \leq S$.

These criteria provide a sufficient condition for the Bit-Interleaved Coded Multiple Beamforming (BICMB) system of [1] to achieve full diversity as the subsequent analysis in [1] shows by replacing $d_{free}$ with the Hamming distance $d_H$ of that errored codeword with the all-zeros codeword.

The interleaver design criteria in Section III of [1] should, similarly, be

For all error paths of interest,
1) the consecutive coded bits are mapped over different symbols,
2) transmitted over different subcarriers of an OFDM symbol,
3) $\alpha_s \geq 1$ for $1 \leq s \leq S$.

For both interleaver design criteria above, if the condition $\alpha_s \geq 1$ is violated for any $s$, $1 \leq s \leq S$, then that particular error pattern can dominate and the full diversity of the resulting BICMB system is not guaranteed.

It can be shown that for the 2x2 BICMB system with the industry standard (133, 171), code rate $R=1/2$, constraint length $K=7$ convolutional code, employing $S = 2$ streams, it is possible to come up with an interleaver that satisfies the interleaver design criteria above for all possible error patterns.

[1] Atheros Communications, Inc., Santa Clara, CA 95054-3644.

[2] Department of EECS, University of California, CA 92697-2625.